\documentclass[11pt]{article}
\usepackage{vietnam,epsfig}

\def\Journal#1#2#3#4{{#1} {\bf #2}, #3 (#4)}

\def\ApJS{{\em ApJ Suppl.}}
\def\ApJ{{\em ApJ}}
\def\ApP{{\em Astropart. Phys.}}
\def\MNRAS{{\em MNRAS}}
\def\NAR{{\em New Astronomy Review}}
\def\AA{{\em Astronomy and Astrophysics}}

\def\be{\begin{equation}}
\def\ee{\end{equation}}
\def\bea{\begin{eqnarray}}
\def\eea{\end{eqnarray}}

\begin{document}
\vspace*{4cm}
\title{INDIRECT MANIFESTATION OF GRBs IN THE GALAXY:\\ COMPTON TRAILS AND POSITRONS}

\author{ETIENNE PARIZOT }

\address{Institut de Physique Nucl\'eaire d'Orsay \\ IN2P3-CNRS/Universit\'e Paris-Sud, 91406 Orsay Cedex, France}

\maketitle\abstracts{Two independent indirect manifestations of Galactic gamma-ray bursts (GRBs) are considered, whose importance for gamma-ray astronomy will depend on a still poorly known -- though astrophysically important -- parameter, namely the GRB repetition timescale in a Galaxy like ours. Both phenomena are expected to lead to observational constrains about this timescale. The first one relates to the Compton scattering of the GRB gamma-rays in the disk of the Galaxy, and the second one to the long-term annihilation pattern of positrons produced just ahead of the ultra-relativistic plasma, by photo-pair production.}

\section{Introduction}

There is an ongoing debate in the gamma-ray burst (GRB) community, concerning the typical timescale $\Delta t_{\mathrm{GRB}}$ between two successive GRBs in a galaxy like ours. If one knew for sure the progenitors and which astrophysical mechanism actually produced the GRBs, this timescale could be estimated theoretically. For the time being, however, it goes the other way round: one tries to constrain $\Delta t_{\mathrm{GRB}}$ observationally to gain an insight into the mechanism at work.

The rate of \emph{observable} GRBs, as derived from the BATSE catalogue,\cite{Paciesas+99} is of the order of $10^{-7}\,\mathrm{yr}^{-1}$ per galaxy: we would see a Galactic GRB every 10~Myr or so. Taking into account a beaming factor of 50--500~\cite{Frail+01,PanKum01}, this would translate into an actual GRB rate of one every 20--200~kyr. If the jets are structured~\cite{Rossi+02} rather than homogeneous, 3--10 times more GRBs could be observed, leaving the actual rate around one every $10^5$--$10^6$~yr~\cite{Podsiadlowski+04}. On the other hand, Wick et al.~\cite{Wick+04} argue for a higher frequency, with $\Delta t_{\mathrm{GRB}}$ around 3--10~kyr. A further complication is that the GRB rate of course depends on what we call a ``GRB''! Analysing the HETE-2 data, Lamb \emph{et al.}~\cite{Lamb+04} recently proposed a uniform jet model unifying X-ray flashes and GRBs. It implies less energetic, but on the other hand more frequent GRBs. The repetition timescale could thus be as low as $10^3$~years or even less, i.e. roughly similar to that of the Type-Ic core collapse supernov\ae.

As can be seen, $\Delta t_{\mathrm{GRB}}$ remains a very uncertain, though crucial parameter, with a value probably somewhere between $10^3$ and $10^6$~yr. We discuss below two astrophysical processes which should make Galactic GRBs indirectly visible long after their explosion, independently of their beam direction, so that a constraint on their actual rate could be derived.

\section{The Compton trails of GRBs}

We first discuss the generation of a so-called \emph{Compton trail} along the path of the GRB photons on their way out of the Galaxy.\footnote{Note that these have nothing to do with the ``Compton tails'' discussed by Barbiellini elsewhere in this volume.} The idea is that the $\gamma$-rays emitted during the burst are in principle subject to Compton scattering as they interact with the ambient electrons (either free or bound to a nucleus in the interstellar medium), and can thus be re-emitted in all directions, to be observed from anywhere in the Galaxy long after their emission from the GRB. 

Let us consider a ``typical GRB'' emitting photons with a mean energy of $E_{\gamma} =
200$~keV, for a total of $E_{\mathrm{GRB}} = E_{51}\times 10^{51}$~erg.  This
corresponds to a huge number of photons, of the order of $N_{\gamma}
\sim 3\,10^{57}\,E_{51}$!  If $n_{\mathrm{e}} = n_{0}\times 1\,\mathrm{cm}^{-3}$ is the typical electron density in the ISM, an estimate of the Compton-scattered flux from a GRB exploding at a distance $D$ is then:
\begin{equation}
    \phi \simeq \frac{N_{\gamma}n_{\mathrm{e}}\sigma_{\mathrm{T}} c}{4\pi D^{2}}
    \approx (0.52\,\mathrm{ph\,cm^{-2}\,s^{-1}}) \times E_{51} n_0
    D^{-2}_{\mathrm{kpc}}.
    \label{eq:phi}
\end{equation}

For Galactic GRBs, with $D_{\mathrm{kpc}}\sim 3$--15~kpc, say, such a flux is far above the detection thresholds of $\gamma$-ray satellites.  It should also be noted that for $n_{\mathrm{e}} = 1\,\mathrm{cm}^{-3}$, the Compton depth of the Galaxy is $\simeq 2\times 10^{-3}$ per kpc, which means that the fraction of GRB photons which are Compton scattered is always small, even if they propagate over several kiloparsecs along the Galactic plane.  For this reason, we can neglect multiple scattering and consider that the primary $\gamma$-ray beam is not affected at all by the process.

\begin{figure}
\centering
\includegraphics[width=0.5\linewidth]{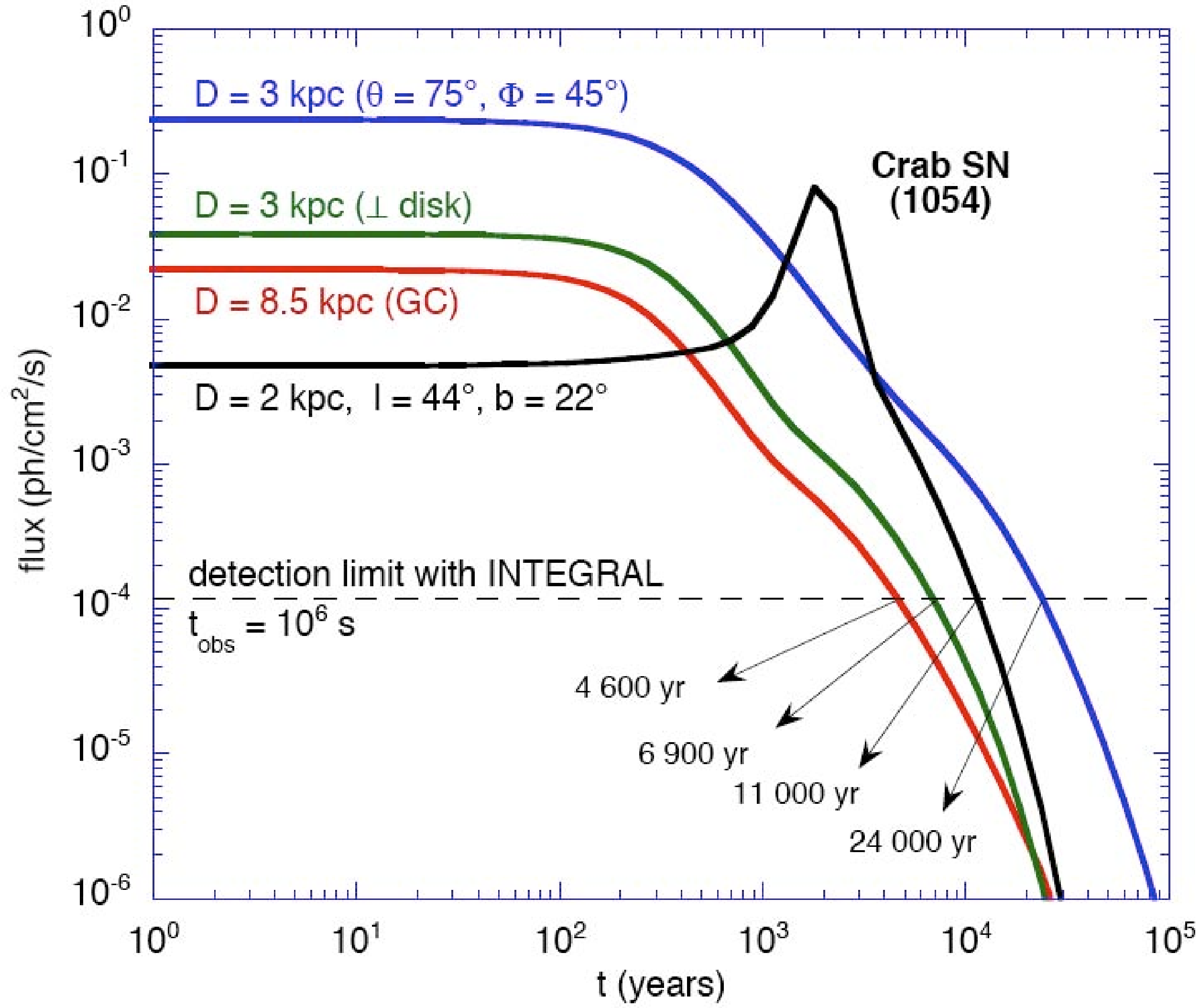}\\
\caption{Light curves of GRB Compton trails, for various GRB distances and orientations in the Galaxy.}
\label{fig:LightCurves}
\end{figure}

The above indirect manifestation of the GRBs is what we call a \emph{Compton trail}. While the prompt emission is observable during at most a hundred second or so -- and only if the observer is inside the emission cone! -- the Compton trail can be seen from anywhere in the Galaxy, all along the journey of the photons out of the disk. The scattered flux essentially drops simply when there is no target electrons anymore. For a disk thickness of 300~pc, say, it will take roughly 1000 years if the emission cone is perpendicular to the disk, 2 times more if it is inclined by 60$^{\circ}$, and up to $10^5$ years for GRB axes within the disk. More detailed 3D calculations including a realistic model of the gas distribution in the Galaxy and of the GRB primary spectrum can be found in our paper.\cite{AllPar04} A few examples of the expected light curves are shown in Fig.~\ref{fig:LightCurves}.

\begin{figure}
\centering
\includegraphics[width=0.8\linewidth]{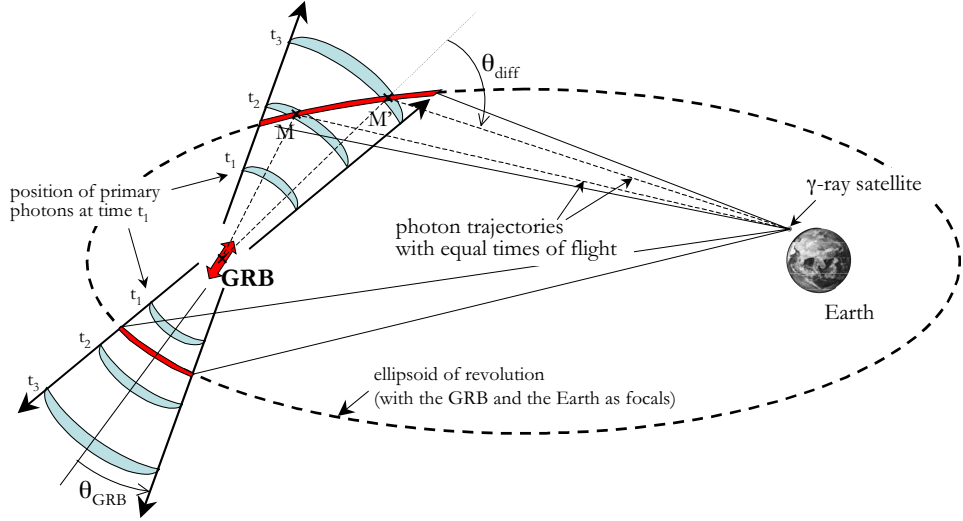}\\
\caption{A schematic view of the Compton trail geometry, as discussed in the text.}
\label{fig:CTSketch}
\end{figure}

A simple geometrical argument allows one to determine the shape of the emission region at any given time, $t$, after the explosion: it is the intersection of the GRB emission cone and the ellipsoid of revolution defined by $GM + ME = ct$, the foci of which are the GRB central object, G, and the Earth, E (see Fig.~\ref{fig:CTSketch}). As seen from Earth, this shows up as a filled ellipse on the sky (with intensity contrasts reflecting directly the local ISM density structure). We therefore suggest that such ellipses should be looked for in the $\gamma$-ray data. If one is found, then one should also look for its counterpart (from the other half of the GRB cone) and then locate geometrically the GRB remnant for further studies. Note that if no such Compton trail ellipse can be found, a lower limit on the time passed since the last GRB explosion in our Galaxy may be derived, and hence a constraint on $\Delta t_{\mathrm{GRB}}$. Given the typical light curves on Fig.~\ref{fig:LightCurves}, the Compton trails appear to be very interesting to explore low values of this parameter, up to $10^4$~yr or so.

\section{Positrons from GRBs and the 511~keV emission from the Galactic bulge}

Another indirect manifestation of GRBs is the annihilation of positrons produced just ahead of the relativistic fireball, when primary photons which are Compton backscattered by the ionized medium upstream interact with subsequent GRB photons via $\gamma\gamma$ pair-production interactions.\cite{ThoMad00,DerBoe00} Assuming a rather conservative conversion efficiency, $\xi_{\mathrm{pair}} = 1$\%,\cite{DerBoe00} the total number of positrons produced is $N_{+}\sim 6\,10^{54}\,E_{51}\,(\xi_{\mathrm{pair}}/0.01)$. In a more detailed study of the spectral modifications of GRBs by pair precursors, M\'esz\'aros et al.\cite{Meszaros+01} calculate a positron yield of $N_{+}\sim 3\,10^{55}E_{51}$, corresponding to $\xi_{\mathrm{pair}}\sim 5$\%.

Furlanetto and Loeb\cite{FurLoe02} have addressed the question of the annihilation of such positrons associated with a GRB exploding in a dense medium (molecular cloud), with $n = n_{30}\times 30\,\mathrm{cm}^{-3}$. They have shown that most of the annihilation signal would arise in the radiation phase of the associated supernova remnant expansion, when a dense shell forms. This would lead to an observable signal lasting for $\tau_{\mathrm{ann}}\sim 10^{4}\, n_{30}^{-4/7}$~yr. If $\Delta t_{\mathrm{GRB}}$ is smaller than this, such a signal should typically be observed in one or a few high mass star forming regions, i.e. in coincidence with superbubbles and/or OB associations.

In a recent paper,\cite{Parizot+05} we have studied the phenomenology of the annihilation of positrons associated with GRBs occurring around the Galactic center, and showed that it should be quite different. There, indeed, the formation of massive stars mostly occurs during phases of mini starbursts\cite{BlaCoh03}. GRBs in the Galactic center are thus likely to explode in a superbubble-like environment in which we showed\cite{Parizot+04} that the associated supernova shock dies \emph{before} becoming radiative. The positrons thus cannot annihilate in the never-formed shell, and are free to diffuse away and fill the Bulge where they annihilate on timescales of $\sim 10^7$~yr. Since this is longer than the repetition timescale of GRBs and of mini starbursts, a steady-state annihilation signal from the whole Galactic bulge is expected.

We have shown in our paper\cite{Parizot+05} that this process can match very naturally the phenomenology of the well-known, but still mysterious 511~keV emission from the Galactic bugle\cite{Paul04,Weidenspointner+04}, provided that the mean time between GRBs occurring \emph{in the bulge} if of $\Delta t_{\mathrm{GRB}} \simeq 8\,10^4\mathrm{yr}\,E_{51}\,(\xi_{\mathrm{pair}}/5\%)$. This can be seen as an interesting constraint on the GRB repetition time (keeping in mind that the Galactic rate is probably 5--10 times larger than the bulge rate), as well as on the GRB-induced positron production mechanism.

It has been suggested recently that the supernova associated with remnant W49B was in fact a GRB, which occurred 3000~years ago\cite{Ioka+04,Keohane+04}. The possibility of detecting the associated e$^{+}$e$^{-}$ annihilation and/or Compton trail will be investigated in a forthcoming paper.

\section*{Acknowledgments}
I wish to thank the organisers for their invitation, Michel Cass\'e, Jacques Paul and Roland Lehoucq for their contribution to the 511~keV emission model of the Galactic bulge, Denis Allard for his help with the Compton trail calculations, Sergio Assad for inspiring me the smile on the front page picture, and Angela Olinto for her numerical eye always at the ready ;-)

\end{document}